\documentclass{ifacconf}

\usepackage{graphicx}      
\usepackage{natbib}   
\usepackage{siunitx}
\usepackage{amsmath}
\usepackage{amssymb}
\usepackage{float}
\usepackage{svg}
\newtheorem{Theorem}{\bf Theorem}

\newtheorem{Remark}[Theorem]{\bf Remark}

\usepackage[ruled,vlined,algo2e]{algorithm2e}
\usepackage{xcolor}

\begin{document}
\begin{frontmatter}

\title{Stochastic MPC for energy hubs using data driven demand forecasting}

\author[First]{Francesco Micheli \,\thanksref{footnoteinfo}} 
\author[First,Second]{Varsha Behrunani\, \thanksref{footnoteinfo}}
\author[First]{Jonas Mehr}
\author[Second]{Philipp Heer} 
\author[First]{John Lygeros}

\thanks[footnoteinfo]{These authors contributed equally to the work.}

\address[First]{Automatic Control Lab, ETH, Zürich 8092, Switzerland (e-mail: mehrjo@student.ethz.ch, {bvarsha, frmicheli, jlygeros}@ethz.ch)}
\address[Second]{Urban Energy Systems Laboratory, Empa, Dübendorf 8600, Switzerland (e-mail:  {varsha.behrunani,philipp.heer}@empa.ch)}

\thanks[footnoteinfo1]{This research was supported by the Swiss National Science Foundation under NCCR
Automation, grant agreement 51NF40\_180545. }
\begin{abstract}
Energy hubs convert and distribute energy resources by combining different energy inputs through multiple conversion and storage components. The optimal operation of the energy hub exploits its flexibility to increase the energy efficiency and reduce the operational costs. However, uncertainties in the demand present challenges to energy hub optimization. In this paper, we propose a stochastic MPC controller to minimize energy costs using chance constraints for the uncertain electricity and thermal demands. Historical data is used to build a demand prediction model based on Gaussian processes to generate a forecast of the future electricity and heat demands. The stochastic optimization problem is solved via the Scenario Approach by sampling multi-step demand trajectories from the derived prediction model. The performance of the proposed predictor and of the stochastic controller is verified on a simulated energy hub model and demand data from a real building.
\end{abstract}


\end{frontmatter}
\section{Introduction}
Energy hubs provide a flexible framework for the supply of several energy types through the conversion and storage of multiple energy sources. Energy hubs are a promising solution to improve energy efficiency and increase the renewable energy penetration by enabling versatile operating strategies. Optimizing the energy hub to achieve an economic goal can be formulated as a constrained optimization.
However, building energy consumption is governed by several complex interactions such as weather, occupancy, thermal properties etc., which make accurate prediction of future energy demand very difficult. Uncertainty in electrical and thermal demands is therefore one of the main challenges for energy hub control.

Several data driven techniques have been proposed in the literature for building energy demand forecasts, and an overview of different models is presented in  \cite{Ahmad:2018, Amasyali:2018}.  In \cite{Kato:2008} and \cite{Akarslan:2018}, ANN's are used for heat load prediction and electrical demand forecasting, respectively. \cite{Bunning:2020} further incorporates online error autocorrelation into ANN's to reduce the variance of thermal load prediction. GP's have previously been used in \cite{ZENG:2020, Fang:2021, lourenco:2012} for the day ahead demand forecast of buildings. \cite{Prakash:2018} incorporates physical insights about load data as prior knowledge into the GP model to improve accuracy and reduce training requirements and \cite{lourenco:2012} studies the impact of different inputs on the model.

The uncertainty in the prediction is incorporated into energy management using stochastic approaches. Chance constrained has been explored for energy hubs in \cite{Hou:2020, Da:2019}. In \cite{Da:2019}, these constraints model the uncertainties in the power and gas flows between hubs and \cite{Cao:2022} uses an ambiguity set along with CVaR approximation for a robust day ahead scheduling method. 
Forecasting using GP's has been used with MPC in \cite{Lee:2018, Zhang:2018, Zhang:2013}. In \cite{Zhang:2013}, a randomized MPC incorporates the weather uncertainty and occupancy data using sampled scenarios, whihc is further extended in \cite{Zhang:2018} using typical scenarios through scenario reduction technique.

In this study, we develop a Gaussian processes (GP) model to forecast electricity and thermal demands of an individual building. We propose a stochastic MPC controller to account for the uncertainty in future energy demands and solve the finite-horizon stochastic optimization problem via the Scenario Approach. This method leverages multi-step demand trajectory samples from the derived prediction model to derive a deterministic sampled version of the original chance constrained finite-horizon optimal control problem. Finally, we numerically evaluate the proposed multi-step prediction algorithm and the scenario-based MPC approach on a simulated energy hub model and demand data from a real building.

\section{Energy Hub Modelling}
\begin{figure}
\begin{center}
\includegraphics[width=8.4cm]{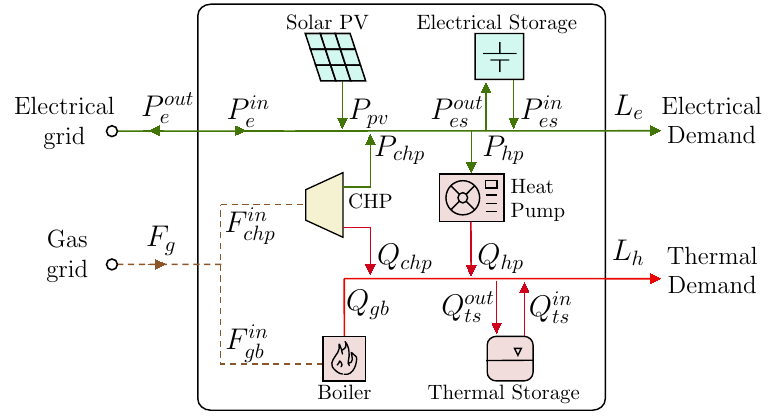}
\caption{Topology of the energy hub} 
\label{fig:ehub}
\end{center}
\end{figure}

The role of the energy hub is to satisfy the uncertain energy demand by combining different energy sources through its energy conversion and storage components. The objective is to minimize the economic cost by exploiting the flexibility stemming from the combination of multiple energy inputs, conversion and storage components. This can be accomplished by optimally scheduling the generation and storage units to satisfy the future energy demand. Therefore, an accurate forecast of the demand is of vital importance to improve the overall energy hub performance.
An example of the topology of an energy hub and the energy flow variables are illustrated in Fig.~\ref{fig:ehub}. The inputs to the energy hub are electricity and natural gas from the grid and the outputs are the electricity and thermal demands. The demands are considered uncontrollable, i.e., they do not coordinate with the energy hub and act as an exogenous disturbance on the system.  
The energy hub comprises a photovoltaic (PV) generation unit, a combined heat and power unit (CHP), a heat pump (HP), a gas boiler (GB), an electrical storage (ES), and a thermal storage (TS). An overview of the model and constraints of each components is given below. The constraints must hold for each time-step $k$. The CHP and the HP couple the electricity and thermal systems as the CHP simultaneously produces electricity and heat using natural gas, while the heat pump uses electricity to produce thermal energy. 

The CHP operation is limited within a feasibility region described by a polyhedron with vertices A, B, C, D, ~\cite{navarro:2018}. The output of the CHP, $P_{\text{chp}}$ and $Q_{\text{chp}}$, is a convex combination of the electrical and thermal output of the vertices, $P_{\text{A}}$,$\cdots$, $P_{\text{D}}$ and $Q_{\text{A}}$, $\cdots$, $Q_{\text{D}}$, with weights $w_{\text{A,i}}$, $\cdots$, and $w_{\text{D,i}}$, respectively. The gas consumed by the CHP and its fuel efficiency are $F^{in}_{\text{chp,k}}$ and $\eta_{\text{chp}}$ respectively. The resulting outputs and efficiencies can be characterized by the following equations:
\begin{equation}\label{eq:CHP}
\small
         \begin{aligned}
          P_{\text{chp,k}} &= \sum_{j \in S} w_{\text{j,k}} \cdot P_{\text{j,k}}& F^{in}_{\text{chp,k}} &= \frac{1}{\eta_{\text{chp}}}\cdot P_{\text{chp,k}}\\
     Q_{\text{chp,k}} &=\sum_{j \in S} w_{\text{j,k}} \cdot Q_{\text{j,k}} & 1&=\sum_{j \in S} w_\text{j,k} \\
     0 &\leq w_\text{j,k} \leq 1&S &=\left\{\text{A, B, C, D}\right\} \end{aligned}\normalsize
         \end{equation}
The relation between the input and output of the heat pump and the boiler are 
\begin{equation}
\label{eq:HP_GB}
\begin{aligned}
Q_{\text{hp,k}} &= \text{COP} \cdot P_{\text{hp,k}}\ ,\\
Q_{\text{gb,k}} &= \eta_{\text{gb}} \cdot F^{in}_{\text{gb,k}}\ ,
\end{aligned}
\end{equation}
where COP is the coefficient of performance for the pump and $\eta_{gb}$ is the boiler efficiency.
The electrical power output of the PV is a function of the incident solar irradiance $I_{\text{solar}}$ $[\text{kW/m}^2]$, the total area  $a_{\text{pv}}$ $[\text{m}^2]$ of the panel, and the efficiency $\eta_{\text{pv}}$.
\begin{equation}\label{eq:PV}
P_{\text{pv,k}} = \eta_{\text{pv}} \cdot I_{\text{solar,k}} \cdot a_{\text{pv}}\ .
\end{equation}
To predict the future output of the PV, the forecast of the global solar irradiance and therefore, $I_{\text{solar}}$ is assumed to be accurately known. Uncertainties in the PV production forecast can also be introduced in a way similar to the demand uncertainty below, but is not considered here.

The converters are also limited by capacity constraints,
\begin{equation}
\label{eq:cLimits}
\begin{aligned}
P^{\text{min}}_{\text{m}} &\leq P_{\text{m,k}}\leq P^{\text{max}}_{\text{m}} \ \ \ \text{m}\in\{\text{pv, chp}\}\ ,\\
Q^{\text{min}}_{\text{n}} &\leq Q_{\text{n,k}}\leq Q^{\text{max}}_{\text{n}} \ \ \ \text{n}\in\{\text{gb, hp, chp}\}\ .
\end{aligned}
\end{equation}
The electrical grid acts both as a source and as a sink for the hub allowing the electricity load to always be satisfied. To prevent simultaneous import ($P^{\text{in}}_{\text{e,k}}$) and export ($P^{\text{out}}_{\text{e,k}}$), the power exchanges with the grid are subject to the constraint,
\begin{equation}P^{\text{in}}_{\text{e,k}} \cdot P^{\text{out}}_{\text{e,k}} =0 \ .\end{equation}
The dynamics of the electrical and thermal storage units are modelled as discrete-time dynamical systems: 
\begin{subequations}
\label{storages_constraints}
\begin{align}
  P_{\text{es},k+1} &= \gamma_{\text{es}} \cdot P_{\text{es,k}} +  \eta_{\text{es}} \cdot P^{\text{out}}_{\text{es,k}}  - \left(\frac{1}{\eta_{\text{es}}}\right) \cdot P^{\text{in}}_{\text{es,k}} \ ,\\
  P^{\text{in}}_{\text{es,k}} & \cdot P^{\text{out}}_{\text{es,k}} =0 \ ,  \\ 
  P^{\text{min}}_{\text{es}} &\leq P_{\text{es,k}}\leq P^{\text{max}}_{\text{es}}\ .
  \end{align}
\end{subequations}
\vspace{-5mm}
\begin{subequations}
\label{storage_h}
\begin{align}
\label{storage_h_constraints}
 Q_{\text{ts},k+1} &= \gamma_{\text{ts}} \cdot Q_{\text{ts,k}} +  \eta_{\text{ts}} \cdot Q^{\text{out}}_{\text{ts,k}}  - \left(\frac{1}{\eta_{\text{ts}}}\right) \cdot Q^{\text{in}}_{\text{ts,k}} 
 \ ,\\ \label{storagelim_h_binary}
  Q^{\text{in}}_{\text{ts,k}} & \cdot Q^{\text{out}}_{\text{ts,k}} =0 \ , \\ \label{storagelim_h}
  Q^{\text{min}}_{\text{ts}} &\leq Q_{\text{ts,k}}\leq Q^{\text{max}}_{\text{ts}} \ ,
\end{align}
\end{subequations}
where the storage energy levels, $P_{es,k+1}$ and $Q_{ts,k+1}$, at time-step $k+1$, depend on the past levels, on the charging or discharging powers, $P^{\text{out}}_{\text{es,k}}$ and $Q^{\text{out}}_{\text{ts,k}}$, and on the standby and cycle efficiencies, $\gamma_{\text{ts}}$ and $\eta_{\text{ts}}$, $\gamma_{\text{es}}$ and $\eta_{\text{es}}$, respectively. The electrical and thermal storage levels are also required to be within some maximum and minimum levels. Additionally, power cannot be charged and discharged in/from the storage simultaneously. 
Finally, the following electrical and thermal power balances need to be satisfied:
\begin{equation}\label{eq:load_balance2}
\begin{aligned}
L_{\text{e,k}} &\!=\!  P_{\text{pv,k}} \!+\! P_{\text{chp,k}}  \!-\! P_{\text{hp,k}} \!+\!  \left(P^{\text{in}}_{\text{e,k}} \!-\! P^{\text{out}}_{\text{e,k}}\right)  \!+\! \left( P^{\text{in}}_{\text{es,k}}  \!-\! P^{\text{out}}_{\text{es,k}}  \right), \\
L_{\text{h,k}} &= Q_{\text{gb,k}}  + Q_{\text{chp,k}} + Q_{\text{hp,k}}  +\left( Q^{\text{in}}_{\text{ts,k}}  - Q^{\text{out}}_{\text{ts,k}}  \right),
\end{aligned}
\end{equation}
where $L_{\text{e}}$ and $L_{\text{h}}$ are the electrical and thermal load demand on the energy hub respectively.
\section{MPC of Energy Hub}
The optimal scheduling of the energy hub is formulated as a finite-horizon problem in which the operational set points are chosen to minimize the operational costs, while satisfying the electrical and thermal load demands, $L_{\text{e}}$ and $L_{\text{h}}$ over the horizon $\mathcal{T}=\SI{24}{\hour}$ with a sampling resolution of $\SI{1}{\hour}$.
Let $p=\{p_k, \, p_{k+1}, \, \dots, p_{k+\mathcal{T}}\}$ collect all the decision variables of the hub over the horizon, where $p_k = \{P_{\text{pv,k}}$,  $P_{\text{chp,k}}$, $P_{\text{hp,k}}$, $P^{\text{out}}_{\text{e,k}}$, $P^{\text{in}}_{\text{e,k}}$, $ P^{\text{out}}_{\text{es,k}}$, $ P^{\text{in}}_{\text{es,k}}$, $ P_{\text{es,k}}$, $Q_{\text{gb,k}}$, $
Q_{\text{chp,k}}$, $Q_{\text{hp,k}}$, $Q^{\text{out}}_{\text{ts,k}}$, $Q^{\text{in}}_{\text{ts,k}}$, $Q_{\text{ts,k}}$, $F^{\text{in}}_{\text{chp,k}}$, $F^{\text{in}}_{\text{gb,k}}\}$.

The resulting finite-horizon economic dispatch problem can be compactly written as:
\begin{equation}
\label{eq:EDP}
\begin{aligned}
\min_{p} \quad & \underbrace{\sum_{\mathcal{T}}\left(P^{\text{out}}_{\text{e,k}} \!\cdot\! C^{\text{out}}_{\text{e,k}} - P^{\text{in}}_{\text{e,k}} \!\cdot\! C^{\text{in}}_{\text{e,k}} + (F^{\text{in}}_{\text{chp,k}}\!+\!F^{\text{in}}_{\text{gb,k}}) \!\cdot\! C_{\text{g,k}} \right)}_{J(p)}
\\
\text{s.t.} \quad
& \quad \eqref{eq:CHP}-\eqref{eq:load_balance2} \quad \forall \  k=1,\dots,\mathcal{T}\ .
\end{aligned}
\end{equation}

$C^{\text{out}}_{\text{e,k}}$ and $C^{\text{in}}_{\text{e,k}}$ are the prices of buying and selling one unit [$\SI{}{\kilo\watt\hour}$] of electricity from the grid, and $C_{\text{g,k}}$ is the price per unit [$\SI{}{\kilo\watt\hour}$] of gas, at time k. The energy hub controller is implemented using an MPC strategy. At each time-step, the optimization determines the optimal control inputs for the system over the finite-horizon $\mathcal{T}$, but only the first of these inputs is applied to the system; the process is then repeated at the next time step.

Since the future energy demands, $L_{\text{e}}$ and $L_{\text{h}}$, are unknown in advance and stochastic in nature, we formulate the economic dispatch problem as a stochastic optimization problem. In particular, we would like to satisfy the electrical and thermal load balance equations at all times. In order to achieve this, the decision variables of the hub, $p$ is split into $p^s$ and $p^L$.
Let $p^s=\{p^s_k, \, p^s_{k+1}, \, \dots, p^s_{k+\mathcal{T}}\}$ collect the operational set points, over the horizon $\mathcal{T}$, with $p^s_k = \{P_{\text{pv,k}}$,  $P_{\text{chp,k}}$, $P_{\text{hp,k}}$, $ P^{\text{out}}_{\text{es,k}}$, $ P^{\text{in}}_{\text{es,k}}$, $ P_{\text{es,k}}$, $Q_{\text{gb,k}}$, $Q_{\text{chp,k}}$, $Q_{\text{hp,k}}$, $F^{\text{in}}_{\text{chp,k}}$, $F^{\text{in}}_{\text{gb,k}}\}$ and $p^L=\{p^L_k, \, p^L_{k+1}, \, \dots, p^L_{k+\mathcal{T}}\}$ be the set of variables that are completely determined by $p^s$ and by the constraints \eqref{eq:CHP}-\eqref{storagelim_h_binary}, \eqref{eq:load_balance2} with $p^L_k = \{ P^{\text{out}}_{\text{e,k}}$, $ P^{\text{in}}_{\text{e,k}}$ $Q^{\text{out}}_{\text{ts,k}}$, $Q^{\text{in}}_{\text{ts,k}}$, $Q_{\text{ts,k}}\}$.

We consider as control variables the set points $p^s_k$, while we exploit the power grid, $P^{\text{out}}_{\text{e,k}}$ and $P^{\text{in}}_{\text{e,k}}$, and the thermal storage, $Q^{\text{out}}_{\text{ts,k}}$ and $Q^{\text{in}}_{\text{ts,k}}$, to compensate for any uncertainty in the electrical and thermal demands. We will assume no limitations on the energy exchange with the electricity grid, $P^{\text{out}}_{\text{e,k}}$ and $P^{\text{in}}_{\text{e,k}}$, leaving the constraint on the thermal storage as the only constraint that depends on the future realization of the demands. Finally, the chance constrained optimization problem is formulated as follows:
\begin{equation}
\label{eq:EDP_chance}
\begin{aligned}
    \min_{p^s,p^L} \quad & \mathbb{E}^{L_{\text{e}},L_{\text{h}}}\left[ J(p^s,p^L) \right] \\
    \text{s.t.} \quad
        &\begin{aligned} 
            & \eqref{eq:CHP}-\eqref{storagelim_h_binary}, \eqref{eq:load_balance2} \ ,  \\
            &\mathbb{P}^{ L_{\text{e}},L_{\text{h}} } \left[
            \begin{aligned}
                &Q^{\text{min}}_{\text{ts}} \leq Q_{\text{ts,k}}\\
                &Q_{\text{ts,k}}\leq Q^{\text{max}}_{\text{ts}}
            \end{aligned}\right] \geq \alpha \ ,
            \\
            & \forall \  k=1,\dots,\mathcal{T}\ .
        \end{aligned}
\end{aligned}
\end{equation}
The cost is formulated as an expectation and the constraint on the thermal storage as a joint chance constraint, both over the distribution of the demand.
\section{Demand forecasting with GP}\label{sec:GPforecast}
To solve the chance constrained stochastic problem we require a forecast of the distributions of the demands. To this end, we develop one-step prediction models based on Gaussian processes, separately for the the electric and thermal demands. These are then recursively employed to forecast the demands over a longer horizon.

\subsection{Background}
Gaussian process (GP) regression is a non-parametric regression method that models a nonlinear function between an input ${x}$ and the output $y=f(x)$ assuming that, for any finite number of inputs $x$, the outputs $y$ are distributed according to a joint multi-variate Gaussian distribution that depends on the values of $x$.

A GP is fully defined by a mean function $\mu(\cdot)$ and a covariance function $\mathbf{K}(\cdot, \cdot)$. Given a set of training/historical data $\mathcal{D}=(\mathbf{x},\mathbf{y})$, the prior joint distribution of the observed outputs and the output ${y}^*$ for a new test input ${x}^*$ can be written as:
\begin{equation}
\left[\begin{array}{c}
\mathbf{y} \\
y^*
\end{array}\right]\sim\mathcal{N}\left(\left[\begin{array}{c}
\mu(\mathbf{x}) \\
\mu\left(x^*\right)
\end{array}\right],\left[\begin{array}{cc}
\mathbf{K}(\mathbf{x}, \mathbf{x}) & \mathbf{K}\left(x^*,\mathbf{x}, \right)^{\top} \\
\mathbf{K}\left(x^*, \mathbf{x}\right) & k\left(x^*, x^*\right)
\end{array}\right]\right) \ ,
\end{equation}
with
\begin{equation}
\begin{aligned}
    &\mathbf{K}(\mathbf{x}^*, \mathbf{x}^*)=\left[\begin{array}{ccc}
    \mathbf{k}\left(x_1^*, x_1^*\right) & \cdots & \mathbf{k}\left(x_1^*, x_N^*\right) \\
    \vdots & \ddots & \vdots \\
    \mathbf{k}\left(x_N^*, x_1^*\right) & \cdots & \mathbf{k}\left(x_N^*, x_N^*\right)
    \end{array}\right]\ ,\\
    &{K}\left(x^*,\mathbf{x}\right) =\left[  \mathbf{k}\left(x^*, x_1\right), \dots, \mathbf{k}\left(x^*, x_N\right)  \right]\ ,
\end{aligned}
\end{equation}
and $\mathbf{k}\left(\cdot, \cdot \right)$ the so called kernel function.

We condition the GP prior on the training data $\mathcal{D}=(\mathbf{x},\mathbf{y})$ to incorporate the information provided by the data about the underlying unknown function. Thus, the posterior distribution, i.e. the predictive distribution, of the output ${y}^*$ for a test input ${x}^*$ can be written as:
\begin{equation}
\left(y^* | x^*, \mathcal{D} \right) \sim \mathcal{N}\left(m^*, \sigma^{*2}\right) \ ,
\end{equation}
with
\begin{equation*}
\small
\begin{aligned}
&m^*=\mu\left(x^*\right)+{K}\left(x^*, \mathbf{x}\right) \left(\mathbf{K}(\mathbf{x}, \mathbf{x})+\sigma_n^2 I\right)^{-1}(\mathbf{y}-\mu(\mathbf{x})), \\
&\sigma^{*2}=\mathbf{k}\left(x^*, x^*\right)-{K}\left(x^*, \mathbf{x}\right) \left(\mathbf{K}(\mathbf{x}, \mathbf{x})+\sigma_n^2 I\right)^{-1} {K}\left(x^*, \mathbf{x}\right)^{\top},
\end{aligned}
\normalsize
\end{equation*}
and $\sigma_n^2$ is the variance of the output noise.
\subsection{Features selection and kernels}
We derived a separate GP-based predictor for the electricity and for the thermal demand:
\begin{equation}
\begin{aligned}
&\hat{L}_{\text{e,k}} \sim \mathcal{N} \left( \mu(x_{\text{e,k}}), \sigma^2(x_{\text{e,k}})\right)\ ,\\
&\hat{L}_{\text{h,k}} \sim \mathcal{N} \left( \mu(x_{\text{h,k}}), \sigma^2(x_{\text{h,k}})\right)\ .
\end{aligned}
\end{equation}
The predicted value of the electrical and thermal energy demand at time-step $k$, $\hat{L}_{e,k}$ and $\hat{L}_{h,k}$ depend on the input feature vectors, $x_{e,k}$ and $x_{h,k}$, formed by a combination of date and time information, weather information such as ambient temperature and solar irradiation, and past electricity or heating demand data. We determined the feature vector through a backward feature selection procedure considering predictive accuracy and computational complexity. This process begins with a large number of features and iteratively trains and evaluates the model, removing the least relevant features.

For the electricity demand we used the feature vector $$ x_{\text{e,k}}\!=\!\Psi_e(k,T_\text{k},L_{\text{e,k-168}},\cdots, L_{\text{e,k-1}}) \!=\! [t_{\text{k}}, T_\text{k}, \Phi_{\text{e,k}}]\ ,$$
where:
\begin{itemize}
    \item $t_{\text{k}}$ includes sin/cos time encoding with periods of one year, one month and one week, and a binary encoding for workday/holiday;
    \item $T_{\text{k}}$ is the ambient temperature at time-step $k$;
    \item $\Phi_{\text{k}}$ includes the electrical demand of the past 6 hours ($L_{\text{e,k-6}},\dots, L_{\text{e,k-1}}$) along with the $5^{\text{th}}$, $50^{\text{th}}$and $95^{\text{th}}$ quantiles of the demand in the past 7 days ($L_{\text{e,k-168}}$, $\dots$, $ L_{\text{e,k-1}}$).
\end{itemize}
Analogously, the same features are used for the heating demand forecasting with the exception that the incident solar irradiation $I_\text{k}$ and the demand values from the past $12$ hours are included: \begin{equation*}x_{\text{h,k}}\!=\!\Psi_{\!\text{h}\!}(k,T_\text{k},I_\text{k},L_{\text{h,k-168}},\cdots, L_{\text{h,k-1}}) \!=\! [t_{\text{k}}, T_\text{k}, I_\text{k}, \Phi_{\text{h,k}}].
\end{equation*}

The GP-based prediction model uses a combination of a radial basis function (RBF) and a linear kernel.
The RBF kernel is applied on all the input dimensions, while the linear kernel is used only on the inputs where a linear relationship to the output can be reasonably expected.
To improve the accuracy of the predictions and account for seasonal dependencies we developed separate model for each season. The hyperparameters of the kernels are tuned using historical data from the past 3 years. The dataset is updated every 24 hours to ensure that the model always has access to the most recent data without updating the hyperparameters. 

\section{Multi-step prediction and scenario program formulation}
Even though the one-step prediction is normally distributed, the distribution of the multi-step demand forecast is generally non-Gaussian. This makes solving the stochastic problem~\ref{eq:EDP_chance} extremely challenging. Therefore, we rely on a sampling based method, namely the Scenario Approach (SA), that allows to handle stochastic problems with generic distributions as long as i.i.d. samples from such distributions are available. We will now focus on how to obtain the multi-step energy demand samples.

Sampling multi-step trajectories from a GP by directly sampling from the function space is generally intractable as it is an infinite dimensional problem. Thus, we resort to a recursive sampling procedure that iteratively queries the one-step GP predictor, using the previous prediction as an input for the next as shown in  Algorithm~\ref{algorithm_1}. 

The samples obtained in this way are i.i.d. samples from the distribution $\mathbb{P}^{L_{\text{e}},L_{\text{h}}}_{GP}$ that will be consider, form now on, the true distribution of the demand. 

\begin{algorithm2e}[t]\label{algorithm_1}
 	\SetAlgoLined
 	\small
 	\KwIn{one-step GP, $\mathcal{D}$.}
 	\For{i=1,\dots,M}{
     	set $\hat{x}_{\text{e,k}}^{(i)} \leftarrow \Psi_e(k,T_\text{k},L_{\text{e,k-168}},\cdots, L_{\text{e,k-1}})$\\
     	\For{s=1,\dots,T}{
     		Sample $\hat{L}^{(i)}_{\text{e,k+s}}\sim \mathcal{N}\big(\mu( \hat{x}_{\text{e,k+s}}^{(i)}), \sigma^2 ( \hat{x}_{\text{e,k+s}}^{(i)}) \big)$\\
     		$\hat{x}_{\text{e,k+s+1}}^{(i)} \leftarrow \Psi_e\big(\scalebox{0.9}{$k\!+\!s\!+\!1$},T_{\scalebox{0.6}{$\text{k}\!+\!\text{s}\!+\!1$}},L_{\scalebox{0.6}{$\text{e,k}\!-\!168\!+\!\text{s}\!+\!1$}},\dots,\hat{L}^{(i)}_{\scalebox{0.6}{$\text{e,k}\!+\!\text{s}$}}\big)$
     		}
 	}\KwRet{$\big[\hat{L}^{(i)}_{\text{e,k}+1}, \dots, \hat{L}^{(i)}_{\text{e,k+T}} \big]$ for $i=1,\dots , M$}
 	\caption{Sampling $M$ multi-step trajectories}
\end{algorithm2e}

\subsection{The Scenario Program}
\label{sec:senario_reformulation}
The Scenario Approach (SA) is based on the solution of a deterministic sampled version of the original chance-constrained problem called the Scenario Program (SP). The SA then provides probabilistic guarantees on how the feasibility of the optimal solution to the SP generalizes to unseen demand realizations.

The SP is reformulated as:
\begin{equation}
	\label{eq:EDP_scenario} 
	\small
	\begin{aligned}
		\min_{p^s, p^L,\sigma^{+}_k, \sigma^{-}_k} \quad & \frac{1}{M}\sum_{i=1}^{M} \left[ J(p^s,p^{L(i)},p^a) \right] \\
		\text{s.t.} \quad 
		&  \eqref{eq:CHP}-\eqref{eq:cLimits}, \eqref{storages_constraints},\\
		&  \sigma_k^{+},\sigma_k^{-}\geq 0,  \\ 
		\begin{aligned}&\forall i = 1, \dots,M\\ &\forall k=1,\dots,\mathcal{T}\end{aligned}
&		\left[\begin{aligned}& P^{\text{out}(i)}_{\text{e,k}} P^{\text{in}(i)}_{\text{e,k}} =0 \\
		&  Q_{\text{ts},k+1}^{(i)} = \gamma_{\text{ts}}  Q_{\text{ts,k}}^{(i)} +  \eta_{\text{ts}} Q^{\text{out}(i)}_{\text{ts,k}}- \left(\frac{1}{\eta_{\text{ts}}}\right) Q^{\text{in}(i)}_{\text{ts,k}} \\ 
		&  Q^{\text{min}}_{\text{ts}} \leq Q_{\text{ts,k}}^{(i)}\leq Q^{\text{max}}_{\text{ts}}\\ 
		& Q^{\text{in}(i)}_{\text{ts,k}} Q^{\text{out}(i)}_{\text{ts,k}} =0 \\
		&  L_{\text{e,k}}^{(i)} =  P_{\text{pv,k}} + P_{\text{chp,k}}  - P_{\text{hp,k}} +\\&\qquad+ (P^{\text{in}(i)}_{\text{e,k}} - P^{\text{out}(i)}_{\text{e,k}}) + ( P^{\text{in}}_{\text{es,k}}  - P^{\text{out}}_{\text{es,k}}) \\
		&  L_{\text{h,k}}^{(i)}\!=\!Q_{\text{gb,k}}  \!+\! Q_{\text{chp,k}} \!+\! Q_{\text{hp,k}}\!+\!( Q^{\text{in}(i)}_{\text{ts,k}}  \!-\! Q^{\text{out}(i)}_{\text{ts,k}}) \\ 
		& Q^{\text{min}}_{\text{ts}} -\sigma^{-}_k \leq Q_{\text{ts,k}}^{(i)}\\
			&Q_{\text{ts,k}}^{(i)}\leq Q^{\text{max}}_{\text{ts}} + \sigma^{+}_k
	\end{aligned}\right] \\
	 \end{aligned}.
	\normalsize
\end{equation}

The variables in the SP with the superscript $(i)$ are dependent on the realization of the $i^{th}$ scenario, i.e. the multi-step trajectory sample obtained through Algorithm~\ref{algorithm_1}.
The expectation in the cost function has been replaced by the empirical (in-sample) mean over the $M$ scenarios, while the (load balance, thermal storage dynamics, and inequality) constraints are enforced on all the samples $i=1,\dots,M$.
As the SP needs to be feasible for all realization of the uncertain demand (which is, by assumption on the model class, defined on an unbounded support), we add the slack variables $\sigma^{+}_k$ and $\sigma^{-}_k$: $\sigma^{+}_k$ is activated when the thermal storage is full and the demand is lower than the thermal production, $\sigma^{-}_k$ is activated when the storage is empty and the demand surpasses the production. The slack variables are linearly weighted in the cost function.
\subsection{A posteriori guarantees and number of scenarios}
As the problem is non-convex, we need to rely on \textit{a posteriori} guarantees regarding the number of scenarios required to ensure that the optimal solution obtained from the SP is, with high probability, a feasible solution to the original chance-constrained problem~\ref{eq:EDP_chance}. We can formalize this statement with the following theorem.
\begin{Theorem}\label{Thm:scenario_guarantees}
	\textit{\cite{campi2018general}}
	Choose the confidence parameter $\beta\in[0,1]$, and let $\varepsilon: \{1,\dots,M\}\rightarrow[0,1]$ satisfy
	\begin{equation*}
    \begin{aligned}
    &\varepsilon(M)=1 \ ,\\
    &\sum_{s=0}^{M-1}\left(\begin{matrix}
    M \\
    s
    \end{matrix}\right)(1-\varepsilon(s))^{M-s}=\beta ,
    \end{aligned}
    \end{equation*}
    then,
    \begin{equation*}
    \mathbb{P}^M\left\{\mathbb{P}^{L_{\text{e}},L_{\text{h}}}_{GP}\left[
            \begin{aligned}
                &Q^{\text{min}}_{\text{ts}} -\sigma^{-}_k \leq Q_{\text{ts,k}}\\
                &Q_{\text{ts,k}}\leq Q^{\text{max}}_{\text{ts}} + \sigma^{+}_k
            \end{aligned}\right] >\varepsilon\left(s_M^*\right)\right\} \leq \beta\ ,
    \end{equation*}
	with $s_M^*$ the cardinality of the (a posteriori) irreducible support subsample, i.e. the cardinality of the smallest set of scenarios for which the optimal solution to the SP remains unchanged. The result holds also for non-minimal irreducible support sets, that can be computed using greedy algorithm, at the expense of a looser bound.
\end{Theorem}
\begin{Remark}
    Note that, as formulated, the guarantees obtained from the SA apply only to the satisfaction of constraints and not to the achievable open-loop cost. To address this issue, we can resort to an epigraph reformulation of the cost function, which introduces one additional decision variable.
\end{Remark}

\begin{Remark}
Note that the guarantees are with respect to the scenario generating distribution, in this case the distribution of the multi-step demands forecast obtained using the iterative sampling Algorithm~\ref{algorithm_1}.
\end{Remark}

Following~\cite{campi2018general}, a possible choice for $\varepsilon(s)$ is: 
\begin{equation*}
\small
	\varepsilon(s):= \left\{
	\begin{aligned}
	    1 \quad \quad \quad& \text { if } s=M \\
    1-\sqrt[M-s]{\frac{\beta}{M\left(\begin{array}{l}
M \\
s
\end{array}\right)}} & \text { otherwise }
\end{aligned}\right.
\normalsize
\end{equation*}

which satisfies
\begin{equation*}
		\varepsilon(s) \leq \frac{1}{M-s} \log \frac{1}{\beta}+\frac{1}{M-s} \log M\left(\begin{matrix}
M \\
s
\end{matrix}\right)
\end{equation*}
highlighting the logarithmic dependence on the confidence parameter $\beta$.
This allows for evaluating the probability of violation level $\epsilon$ achieved. If this is not sufficient, one could repeat the procedure by sampling a larger number of constraints. However, since we are operating in a receding horizon setting, the MPC's intrinsic robustness enables us to significantly reduce the number of samples required to reach the desired probability of violation level $\epsilon$. The next section presents how the number of scenarios affects the closed-loop performance of the controller.

\section{Numerical simulation} 
The proposed prediction model and controller are tested using demand data from the ETZ building at ETH Z\"urich, Switzerland, and a simulated model of the energy hub.

\begin{figure}
\begin{center}
\includegraphics[width=\linewidth]{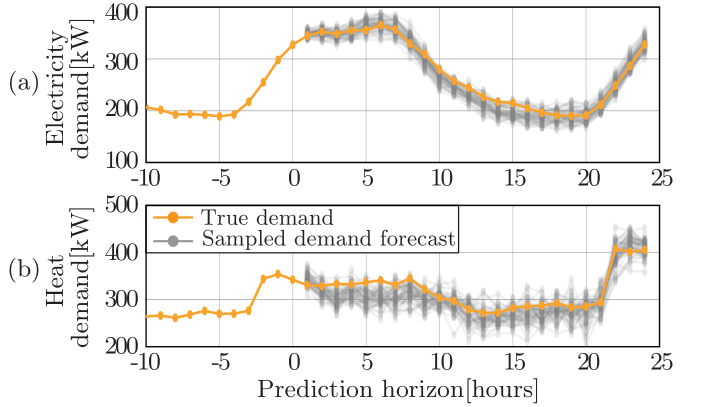}
\caption{Forecasts for (a) electricity, and (b) heat demands on 30 Jan. 2019 at 9:00 AM (hour 0). $50$ scenarios realizations for $\SI{24}{\hour}$ are sampled from the prediction model at time 0.}
\label{fig:trajectories_samples}
\end{center}
\end{figure}

\begin{figure}
\begin{center}
\includegraphics[width=\linewidth]{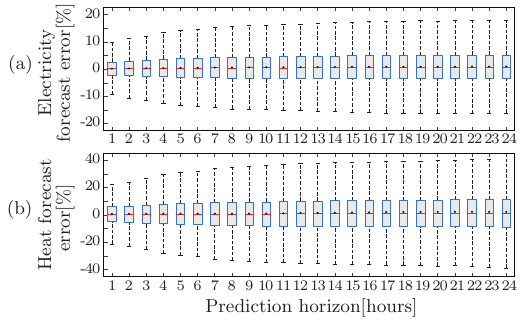}
\caption{Electricity (a) and heat (b) energy demand forecasts error using the multi-step trajectory sampling between 1 Jan. 2019 and 25 Feb. 2019.}
\label{fig:boxplot}
\end{center}
\end{figure}

To evaluate the predictor's performance, we plot the empirical distributions of the residuals over the prediction horizon of $\SI{24}{\hour}$. For each hour between 1 Jan. 2019 and 25 Feb. 2019, we sample 50 multi-step energy demand realizations for $\SI{24}{\hour}$. In Fig.~\ref{fig:trajectories_samples}, we provide an example of how the trajectory samples predict the future electricity and heat demands for $\SI{24}{\hour}$ on 30 Jan. 2019 at 9:00 AM. The error between the predicted demands and the true realizations for different prediction horizons across the considered period is shown in Fig.~\ref{fig:boxplot}. As expected the predictions are more accurate for shorter horizons, while still maintaining the error first and third quartiles between $\pm 5 \%$ for the electricity and $\pm 10 \%$ for the heat forecasts.

\begin{figure}
\begin{center}
\includegraphics[width=8.4cm]{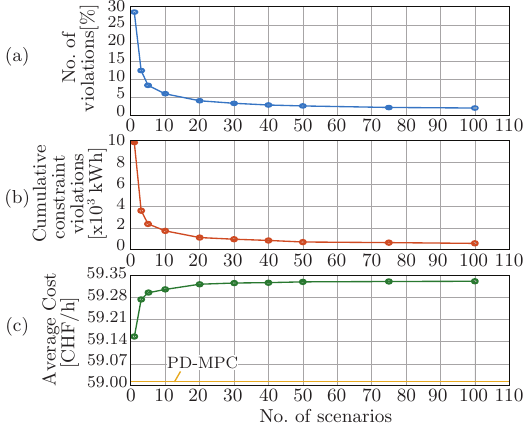}
\caption{Impact of the number of scenarios on the thermal constraint violation and average cost} 
\label{fig:scenario_constraint}
\end{center}
\end{figure}

\begin{figure}
\begin{center}
\includegraphics[width=8.4cm]{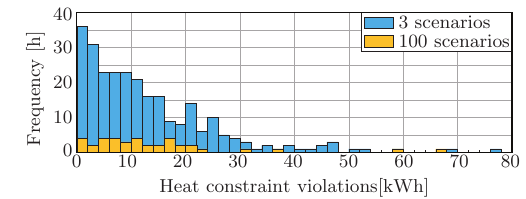}
\caption{Histogram showing the number of thermal constraint violation of different severity in closed loop with the MPC controller for 3 and 100 scenarios}
\label{fig:hist3_100}
\end{center}
\end{figure}

\begin{table}[h]

        \centering
    \centering
    \begin{tabular}{l c c}
                \hline 
            \multicolumn{2}{l}{Component\qquad Parameter} & Value \\ \hline
                CHP & $\eta_{\text{chp}}$, [$P_\text{A}$, $P_\text{B}$, $P_\text{C}$, $P_\text{D}$] &0.36,
               [120, 106, 252, 305]~kW 
                \\
                 & [$Q_\text{A}$, $Q_\text{B}$, $Q_\text{C}$, $Q_\text{D}$]&[0, 171, 408, 0]~kW \\ 
                HP & COP, [$Q_{\text{hp}}^{\text{min}}$, $Q_{hp}^{\text{max}}$] & 4.5, [0, 120]~kW\\ 
                GB  & $\eta_{\text{gb}}$, [$Q_{\text{gb}}^{\text{min}}$, $Q_{\text{gb}}^{\text{max}}$] & 0.78, [0, 120]~kW\\
                PV &$\eta_{\text{pv}}$, $a_{\text{pv}}$, [$P_{\text{pv}}^{\text{min}}$, $P_{\text{pv}}^{\text{max}}$]&0.15, $\SI{3000}{\meter}^2$,  [0, 400]~kW \\
                ES  & $\eta_{\text{es}}$, $\gamma_{\text{es}}$, [$P_{\text{es}}^{\text{min}}$, $P_{\text{es}}^{\text{max}}$]&0.95, 0.999, [40, 250]~kWh\\
            TS  & $\eta_{\text{ts}}$,$\gamma_{\text{ts}}$, l[$Q_{\text{ts}}^{\text{min}}$, $Q_{\text{ts}}^{\text{max}}$]&0.99,0.992,[0, 4800]~kWh\\[1em]
        \end{tabular}
        \caption{Energy hub parameters and capacities}
                \label{tab:ehub_parameters}  
               
\end{table}
\normalsize
The performance of the scenario-based MPC with GP-based forecast is evaluated by simulating the energy hub operation for a period of $3$ months, from 1 Dec. 2018 to 28 Feb. 2019, for a number of scenarios ranging from $1$ to $100$. The energy hub's topology is shown in Fig.~\ref{fig:ehub}, and the model's parameters are summarized in Table~\ref{tab:ehub_parameters}.
In Fig.~\ref{fig:scenario_constraint}, we compare the proposed approach with the MPC controller that has perfect demand information (PD-MPC) in terms of cost and constraint violations. The MPC controller with perfect demand information achieves the lowest average operating cost of 59.01 CHF/h, without any constraint violation. However, this should be interpreted as an unachievable baseline, since assuming perfect demand information is unrealistic.
Increasing the number of scenarios results in a significant reduction of both the number and cumulative constraint violations, at the expense of a small increase in the mean operational cost. Fig.~\ref{fig:hist3_100} shows a comparison between the empirical distributions of the heat constraint violations using $3$ and $100$ scenarios.

\section{Conclusions}
In this paper, we presented a stochastic MPC controller for the optimal dispatch of an energy hub with GP-based demand forecast predictions. The GP-based predictor forecasts the uncertain future energy demands using a relatively limited amount of data. The chance-constrained stochastic MPC problem is solved via the Scenario Approach, using multi-step trajectory samples from the predictor. The proposed approach was tested using real demands data on a simulated energy hub model. The results highlight the role of the number of scenarios by comparing the performance of the scenario-based approach with that of the controller that has perfect demand information in terms of cost and constraint violations. 
Future work aims to extend the proposed framework to a distributed setting with multiple energy hubs operating in a network.

\bibliography{ifacconf}
\end{document}